\documentclass[12pt]{article}
 \usepackage{epsfig}
 \def\be{\begin{equation}}
 \def\ee{\end{equation}}
 \def\bea{\begin{eqnarray}}
 \def\eea{\end{eqnarray}}
 \usepackage{graphicx}% Include figure files

 \catcode`\@=11
 \def\lsim{\mathrel{\mathpalette\@versim<}}
 \def\gsim{\mathrel{\mathpalette\@versim>}}
 \def\@versim#1#2{\vcenter{\offinterlineskip
 \ialign{$\m@th#1\hfil##\hfil$\crcr#2\crcr\sim\crcr } }}
 \catcode`\@=12

 \catcode`@=12
\begin{document}
 \thispagestyle{empty}
 \begin{flushright}
 UCRHEP-T596\\
 Nov 2018\
 \end{flushright}
 \vspace{0.6in}
 \begin{center}
 {\LARGE \bf $U(1)_\chi$ and Seesaw Dirac Neutrinos\\}
 \vspace{1.2in}
 {\bf Ernest Ma\\}
 \vspace{0.2in}
{\sl Physics and Astronomy Department,\\ 
University of California, Riverside, California 92521, USA\\}
%\vspace{0.1in}
%{\sl Jockey Club Institute for Advanced Study,\\ 
%Hong Kong University of Science and Technology, Hong Kong, China\\} 
\end{center}
 \vspace{1.2in}

\begin{abstract}\
In the context of $SO(10) \to SU(5) \times U(1)_\chi$, it is shown how 
seesaw Dirac neutrinos may be obtained.  In this framework, $U(1)$ lepton 
number is conserved, with which self-interacting dark matter with a light 
scalar dilepton mediator may be implemented.  In addition, $U(1)$ 
baryon number may be broken to $(-1)^{3B}$, thereby generating a 
baryon asymmetry of the Universe.  The axionic solution to the strong 
$CP$ problem may also be incorporated.
\end{abstract}

 \newpage
 \baselineskip 24pt

\noindent \underline{\it Introduction}~:~
In considering $SO(10)$ grand unification, the common approach is to allow 
an intermediate step with left-right symmetry, i.e. 
$SU(4) \times SU(2)_L \times SU(2)_R$ or 
$SU(3)_C \times SU(2)_L \times SU(2)_R \times U(1)_{(B-L)/2}$.  This has the 
advantage of forcing into existence the right-handed $SU(2)_R$ lepton doublet 
$(\nu,l)_R$, so that $\nu_R$ is the Dirac partner of the observed $\nu_L$ 
which belongs to the $SU(2)_L$ lepton doublet $(\nu,l)_L$.  At the same time, 
$B-L$ becomes a gauge symmetry, and its breaking through an $SU(2)_R$ scalar 
triplet from the 126 of $SO(10)$ also makes $\nu_R$ massive, realizing thus 
the canonical seesaw mechanism for a naturally small Majorana $\nu_L$ mass. 

Another option~\cite{m18,m19} is to consider 
$SO(10) \to SU(5) \times U(1)_\chi$. 
This is seldom studied because $SU(5)$ is a grand unified symmetry by itself, 
so $U(1)_\chi$ is often thought to be unnecessary and uninteresting. 
Let the breaking of $SU(5)$ to the standard-model (SM) gauge symmetry 
$SU(3)_C \times SU(2)_L \times U(1)_Y$ be at the scale $M_U$.  If 
$U(1)_\chi$ survives down to a scale much below $M_U$ and not too far 
above the electroweak scale, there could be important consequences 
which have been largely overlooked.  In fact, whereas the right-handed 
neutrino $\nu_R$ is a singlet under $SU(5)$, it has a nonzero charge 
under $U(1)_\chi$.  The Higgs doublet which connects $u_L$ to $u_R$ also 
connects $\nu_L$ to $\nu_R$.  Hence a Dirac neutrino mass is again 
obtained and the seesaw mechanism operates as in the left-right case. 
On the other hand, the detailed phenomenology is very different.  Whereas 
$W_R^\pm$ must exist at the left-right scale $M_R$, it must be heavier 
than $M_U$ if $SO(10) \to SU(5) \times U(1)_\chi$.  The $Z_\chi$ gauge boson 
itself has well-defined couplings to the SM particles.  Its existence is 
routinely searched for at the Large Hadron Collider (LHC), with the present 
mass limit~\cite{atlas-chi-17,cms-chi-18} of about 4.1 TeV, which may be 
improved~\cite{kkm18}.

In this paper, new fermions and scalars transforming under $U(1)_\chi$ are 
added to the SM to obtain a number desirable features.  With the help of 
a softly broken $Z_2$ discrete symmetry, natrurally light seesaw Dirac 
neutrinos~\cite{rs84,m02,mpsz15} may be obtained.  The resulting Lagrangian 
conserves both $B$ and $L$.  Further addition of two scalars with $L=-1,-2$ 
enables the appearance of self-interacting leptonic dark matter~\cite{m18-1}. 
The analog of leptogenesis (through a heavy singlet Majorana fermion which 
couples to leptons in the seesaw mechanism) is possible using a heavy singlet 
Majorana fermion which couples to a scalar diquark and an antiquark, 
thereby generating the baryon asymmetry of the Universe.  A fermion color 
octet may also be introduced to support a Peccei-Quinn 
symmetry to obtain an invisible axion for solving the strong $CP$ problem.

\noindent \underline{\it Seesaw Dirac Neutrinos}~:~
The spinorial 16 representation is again chosen for the three families 
of quarks and leptons and their decompositions shown in Table~1.  
\begin{table}[tbh]
\centering
\begin{tabular}{|c|c|c|c|c|c|c|c|}
\hline
fermion & $SO(10)$ & $SU(5)$ & $SU(3)_C$ & $SU(2)_L$ & $U(1)_Y$ & 
$U(1)_\chi$ & $Z_2$ \\
\hline
$d^c$ & 16 & $5^*$ & $3^*$ & 1 & 1/3 & 3 & + \\ 
$(\nu,e)$ & 16 & $5^*$ & 1 & 2 & $-1/2$ & 3 & + \\ 
\hline
$(u,d)$ & 16 & 10 & 3 & 2 & 1/6 & $-1$ & + \\ 
$u^c$ & 16 & 10 & $3^*$ & 1 & $-2/3$ & $-1$ & + \\ 
$e^c$ & 16 & 10 & 1 & 1 & 1 & $-1$ & + \\ 
\hline
$\nu^c$ & 16 & 1 & 1 & 1 & 0 & $-5$ & $-$ \\ 
\hline
$N$ & $126^*$ & 1 & 1 & 1 & 0 & 10 & $-$ \\ 
$N^c$ & 126 & 1 & 1 & 1 & 0 & $-10$ & $-$ \\ 
\hline
\end{tabular}
\caption{Fermion content of model.}
\end{table}
The 
necessary Higgs scalars for fermion masses belong to the 10 representation, 
as shown in Table~2.  
\begin{table}[tbh]
\centering
\begin{tabular}{|c|c|c|c|c|c|c|c|}
\hline
scalar & $SO(10)$ & $SU(5)$ & $SU(3)_C$ & $SU(2)_L$ & $U(1)_Y$ & 
$U(1)_\chi$ & $Z_2$ \\
\hline
$(\phi_1^0,\phi_1^-)$ & 10 & $5^*$ & $1$ & 2 & $-1/2$ & $-2$ & + \\ 
$(\phi_2^+,\phi_2^0)$ & 10 & $5$ & 1 & 2 & $1/2$ & 2 & + \\ 
\hline
$(\eta^+,\eta^0)$ & 144 & 5 & 1 & 2 & 1/2 & $7$ & $-$ \\ 
$\sigma$ & 16 & 1 & $1$ & 1 & $0$ & $-5$ & + \\ 
\hline
\end{tabular}
\caption{Scalar content of model.}
\end{table}

New fermions $N,N^c$ belonging to $126^*,126$ 
respectively are added per family, as well as a Higgs doublet from 144 
and a singlet from 16.  Note that their $Q_\chi$ charges are fixed by the 
$SO(10)$ representations from which they come.  It should also be clear that 
incomplete $SO(10)$ and $SU(5)$ multiplets are considered here (which is the 
case for all realistic grand unified models).  
An important $Z_2$ discrete symmetry is imposed so that 
$\nu^c,N,N^c$ and $\eta$ are odd, and the other fields are even. 
Since $\Phi_1^\dagger$ transforms exactly like $\Phi_2$, the linear 
combination $\Phi = (v_1 \Phi_1^\dagger + v_2 \Phi_2)/\sqrt{v_1^2+v_2^2}$ 
is the analog of the standard-model Higgs doublet, where 
$\langle \phi^0_{1,2} \rangle = v_{1,2}$. 
The $Z_2$ symmetry is respected by all dimension-four terms of the Lagrangian. 
It will be broken softly by the dimension-three trilinear term 
$\mu \sigma \Phi^\dagger \eta$ as well as spontaneously by 
$\langle \eta^0 \rangle = v_3$. 
The $4 \times 4$ neutrino mass matrix spanning $(\nu,\nu^c,N,N^c)$ is 
then given by
\begin{equation}
{\cal M}_\nu = \pmatrix{0 & 0 & 0 & f_\eta v_3 \cr 0 & 0 & f_\sigma u & 0 \cr 
0 & f_\sigma u & 0 & m_N \cr f_\eta v_3 & 0 & m_N & 0},
\end{equation}
where $u = \langle \sigma \rangle$ which breaks $U(1)_\chi$.  The above 
mass matrix generates a seesaw Dirac neutrino with 
$m_\nu = f_\eta f_\sigma v_3 u/m_N$, which is naturally small.

\noindent \underline{\it Scalar Sector}~:~
The scalar potential consisting of $\Phi$, $\eta$, and $\sigma$ is given by
\begin{eqnarray}
V &=& \mu_\Phi^2 \Phi^\dagger \Phi + \mu_\eta^2 \eta^\dagger \eta + \mu_\sigma^2 
\sigma^* \sigma + [\mu \sigma \Phi^\dagger \eta + H.c.] \nonumber \\ 
&+& {1 \over 2} \lambda_\Phi (\Phi^\dagger \Phi)^2 + {1 \over 2} \lambda_\eta 
(\eta^\dagger \eta)^2 + {1 \over 2} \lambda_\sigma (\sigma^* \sigma)^2 \nonumber 
\\ &+& \lambda_{\Phi \eta} (\Phi^\dagger \Phi)(\eta^\dagger \eta) + 
\lambda_{\Phi \sigma} (\Phi^\dagger \Phi)(\sigma^* \sigma) + \lambda_{\eta \sigma} 
(\eta^\dagger \eta)(\sigma^* \sigma).
\end{eqnarray}
The minimum of $V$ satisfies the conditions
\begin{eqnarray}
0 &=& \mu_\Phi^2 + \lambda_\Phi v^2 + \lambda_{\Phi \eta} v_3^2 + \lambda_{\Phi \sigma} 
u^2 + \mu v_3 u/v, \\ 
0 &=& \mu_\eta^2 + \lambda_\eta v_3^2 + \lambda_{\Phi \eta} v^2 + \lambda_{\eta \sigma} 
u^2 + \mu v u/v_3, \\ 
0 &=& \mu_\sigma^2 + \lambda_\sigma u^2 + \lambda_{\Phi \sigma} v^2 + 
\lambda_{\eta \sigma} v_3^2 + \mu v v_3/u. 
\end{eqnarray}
Assuming that $u >> v >> v_3$, the solutions to the above are
\begin{equation}
u^2 \simeq -\mu_\sigma^2/\lambda_\sigma, ~~~ v^2 \simeq -(\mu_\Phi^2 + 
\lambda_{\Phi \sigma} u^2)/\lambda_\Phi, ~~~ v_3 \simeq -\mu u v/(\mu_\eta^2+
\lambda_{\eta \sigma} u^2).
\end{equation}
The $3 \times 3$ mass-squared matrix spanning 
$\sqrt{2}[Im(\phi^0),Im(\eta^0),Im(\sigma)]$ is given by
\begin{equation}
{\cal M}^2_I = -\mu \pmatrix{v_3 u/v & -u & -v_3 \cr -u & vu/v_3 & v 
\cr -v_3 & v & v_3 v/u},
\end{equation}
which has two zero eigenvalues and one massive eigenstate with
\begin{equation}
m^2_{\eta_I} = -\mu \left( {v_3 u \over v} + {v u \over v_3} + {v_3 v \over u} 
\right) \simeq -{\mu v u \over v_3} \simeq \mu_\eta^2 + \lambda_{\eta \sigma} u^2.
\end{equation}
The $3 \times 3$ mass-squared matrix spanning 
$\sqrt{2}[Re(\phi^0),Re(\eta^0),Re(\sigma)]$ is given by
\begin{equation}
{\cal M}^2_R = 2 \pmatrix{\lambda_\phi v^2 & \lambda_{\Phi \eta} v_3 v & 
\lambda_{\Phi \sigma} vu \cr \lambda_{\Phi \eta} v_3 v & \lambda_\eta v_3^2 & 
\lambda_{\eta \sigma} v_3 u \cr \lambda_{\Phi \sigma} v u & \lambda_{\eta \sigma} 
v_3 u & \lambda_\sigma u^2} -\mu \pmatrix{v_3 u/v & -u & -v_3 \cr -u & vu/v_3 
& -v \cr -v_3 & -v & v_3 v/u},
\end{equation}
which is approximately diagonal with
\begin{equation}
m^2_{\phi_R} \simeq 2\lambda_\Phi v^2, ~~~ m^2_{\eta_R} \simeq m^2_{\eta_I}, ~~~ 
m^2_{\sigma_R} \simeq 2 \lambda_\sigma u^2.
\end{equation}
Note that the $\phi_R - \sigma_R$ mixing (with the assumption that $\phi_R$ 
is much lighter than $\sigma_R$) is roughly 
$\lambda_{\Phi \sigma} v u / \lambda_\sigma u^2$ which is naturally suppressed 
by $v/u$.  As for $\eta_R$, its mass is dominated by $-\mu v u/v_3$, and its 
mixing with $\phi_R$ and $\sigma_R$ is suppressed by $v_3/v$ and $v_3/u$ 
respectively.  This justifies the diagonal approximation assumed here.

As a numerical example, let $u=10$ TeV, $v_3=10$ GeV, then $m_\nu = 0.1$ eV 
is obtained for $f_\eta = f_\sigma = 0.1$ and $m_N = 10^{13}$ GeV. 
The soft $Z_2$ breaking parameter $\mu$ is then 6 GeV for $m_{\eta_I} = 1$ TeV.

\noindent \underline{\it Conserved Baryon and Lepton Numbers with 
Self-Interacting Dark Matter}~:~
Because of the $U(1)_\chi$ assignments of the particle content of this model, 
the resulting Lagrangian conserves both baryon number $B$ and lepton 
number $L$ even after the symmetry breaking of $U(1)_\chi$ by $u,v,v_3$ 
and $SU(2)_L \times U(1)_Y$ by $v,v_3$.  As usual, the quarks have 
$B=1/3,L=0$ and the leptons (including the Dirac neutrinos) have $B=0,L=1$.  
This means that if a new particle is added, it may be assigned $B$ and $L$ 
numbers appropriately, according to its assumed interactions with the known 
quarks and leptons~\cite{m15,m17}.  These assignments lie outside 
$U(1)_\chi$, hence $Q_\chi$ 
is now not a marker of dark matter, as in previous studies~\cite{m18,m19}.

\begin{table}[tbh]
\centering
\begin{tabular}{|c|c|c|c|c|c|c|c|}
\hline
scalar & $SO(10)$ & $SU(5)$ & $SU(3)_C$ & $SU(2)_L$ & $U(1)_Y$ & 
$U(1)_\chi$ & $L$ \\
\hline
$\zeta$ & 126 & $1$ & $1$ & 1 & $0$ & $-10$ & $-2$ \\ 
\hline
$\rho$ & 16 & $1$ & 1 & 1 & $0$ & $-5$ & $-1$ \\ 
\hline
\end{tabular}
\caption{Leptonic scalars for self-interacting dark matter.}
\end{table}

For example, consider the scalar singlet $\zeta \sim (1,-10)$ from the 
\underline{126} of $SO(10)$.  It has the allowed Yukawa coupling 
$\zeta^* \nu^c \nu^c$.  In conventional models, $\zeta$ is assumed to have a 
vacuum expectation value, thereby breaking $U(1)_\chi$ and giving $\nu^c$ 
a large Majorana mass, breaking thus also lepton number $L$.  Here it 
may be assumed instead that $L$ is conserved and $\zeta$ has $L=-2$. 
Note that $U(1)_\chi$ is broken instead by $\sigma$ which may be assigned 
$L=0$, together with $L=\pm 1$ for $N,N^c$, and $L=0$ for $\Phi$ and $\eta$.

With $\zeta$ as a scalar dilepton which couples only to the Dirac neutrinos, 
it is then a simple step to consider a scalar singlet $\rho \sim (1,-5)$ 
from the \underline{16} of $SO(10)$ with $L=-1$ so that it can be  
self-interacting dark-matter~\cite{kkpy17} with $\zeta$ as its light mediator, 
as proposed recently~\cite{m18-1}.  Since $\zeta$ decays only to two 
neutrinos, it does not disrupt the cosmic microwave background (CMB) from 
its enhanced production at late times due to the Sommerfeld effect.  It 
removes an important objection~\cite{gibm09,bksw17} to models where the 
light mediator decays to electrons and photons, usually through Higgs 
mixing, which is forbidden here by $L$ conservation.

\noindent \underline{\it Baryogenesis}~:~
Since lepton number is strictly conserved, the usual mechanism of 
generating the baryon asymmetry of the Universe through leptogenesis is 
not possible.  However, the analog process of having a heavy Majorana 
fermion $\psi$ decaying to $B = \pm1$ final states~\cite{m88} may be 
implemented with the addition of two scalar diquarks $h_{1,2}$, as shown 
in Table 4.

\begin{table}[tbh]
\centering
\begin{tabular}{|c|c|c|c|c|c|c|c|}
\hline
scalar/fermion & $SO(10)$ & $SU(5)$ & $SU(3)_C$ & $SU(2)_L$ & $U(1)_Y$ & 
$U(1)_\chi$ & $B$ \\
\hline
$h_1$ (scalar) & $16^*$ & $5$ & 3 & $1$ & $-1/3$ & $-3$ & $-2/3$ \\ 
$h_2$ (scalar) & $10$ & $5$ & 3 & $1$ & $-1/3$ & $2$ & $-2/3$ \\ 
\hline
$\psi$ (fermion) & 45 & $24$ & $1$ & 1 & $0$ & $0$ & $1$ \\ 
\hline
\end{tabular}
\caption{New particles for baryogenesis.}
\end{table}

The allowed couplings involving the new particles are
\begin{equation}
h_2 u d, ~~~ h_2^* u^c d^c, ~~~ h_1 d^c \psi, ~~~ h_1^* h_2 \sigma.
\end{equation}
Assuming a large Majorana mass for $\psi$ which breaks 
$B$ to $(-1)^{3B}$, it may now decay 
to both $h_1^* \bar{d}^c$ ($B=1$) and $h_1 d^c$ ($B=-1$).  The subsequent 
decay of $h_1$ to $u^cd^c$ through $h_1-h_2$ mixing from 
$\langle \sigma \rangle$ establishes a baryon asymmetry in analogy to 
the case of a lepton asymmetry if $\nu^c$ 
were a heavy Majorana fermion.  The one-loop vertex~\cite{fy86} and 
self-energy~\cite{fps95}  
diagrams which contribute to the $CP$ asymmetry and thus the $B$ asymmetry 
are depicted in Fig.~1.
\begin{figure}[htb]
\vspace*{-6cm}
\hspace*{-3cm}
\includegraphics[scale=1.0]{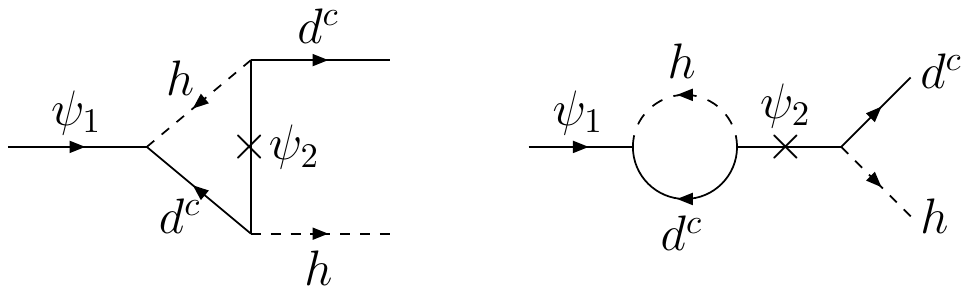}
\vspace*{-21.5cm}
\caption{One-loop diagrams for baryogenesis.}
\end{figure}
They are completely analogous to those of leptogenesis where $\psi_{1,2}$ 
are replaced by $\nu^c_{1,2}$, the scalar diquark $h$ by $\phi^+$, and 
$d^c$ by $l$. 
However, since the Yukawa interactions $h d^c \phi_{1,2}$ are not 
constrained by lepton masses as in $\phi^+ l \nu^c_{1,2}$, the desirable 
asymmetry is easily obtained.  Note of course that 
$m_h << m_{\psi_1} <m_{\psi_2}$ is assumed.  The resulting $B$ asymmetry 
is converted to a $B-L$ asymmetry through the spheralons, again in 
analogy to what happens in leptogenesis.

A possible scenario is high-scale baryogenesis with $m_{\psi_1} \sim 10^{13}$ 
GeV.  Instead of three families of leptons in leptogenesis, just one set of 
scalar diquarks $h_{1,2}$ is needed.  The $CP$ asymmetry generated by the decay 
of $\psi_1$ assuming that $\psi_2$ is much heavier is given by
\begin{equation}
\epsilon = -{3 \over 16 \pi} {m_{\psi_1} \over m_{\psi_2}} {Im[(f_1f_2^*)^2] \over 
|f_1|^2},
\end{equation}
where the $\psi_1$ decay rate is $\Gamma_1 = |f_1|^2 m_{\psi_1}/8 \pi$.  
Consider the parameter $K = \Gamma_1/H(T=m_{\psi_1})$, where the Hubble 
parameter is $H=1.66 \sqrt{g_*}(T^2/M_{Pl})$, as a measure of the deviation 
from equilibrium.  If $K << 1$, which means $|f_1| << 0.02$, then the baryon 
asymmetry is of order $\epsilon/g_*$.  Setting this to $10^{-10}$, and 
assuming $m_{\psi_2}/m_{\psi_i} = 6$, then $|f_2| = 10^{-3}$ if the relative 
phase between $f_1$ and $f_2$ is of order 1.

\noindent \underline{\it Axionic dark matter}~:~
To obtain an axionic solution to the strong $CP$ problem, a colored fermion 
is needed which has an anomalous Peccei-Quinn charge.  Instead of the usual 
quark triplet, a fermion color octet, such as the gluino of supersymmetry, 
may be used~\cite{dm00}.  In a nonsupersymmetric context, it may just be 
any fermion color octet~\cite{mot17} unrelated to the gluon.  Here it is 
\begin{table}[tbh]
\centering
\begin{tabular}{|c|c|c|c|c|c|c|c|}
\hline
scalar/fermion & $SO(10)$ & $SU(5)$ & $SU(3)_C$ & $SU(2)_L$ & $U(1)_Y$ & 
$U(1)_\chi$ & $PQ$ \\
\hline
$\Omega$ (fermion) & 45 & $24$ & $8$ & 1 & $0$ & $0$ & 1 \\ 
\hline
$S$ (scalar) & 54 & $24$ & $1$ & 1 & $0$ & $0$ & 2 \\ 
\hline
\end{tabular}
\caption{New particles for strong $CP$ conservation through the axion.}
\end{table}
called $\Omega$ and it obtains a large Majorana mass through the interaction 
$S^* \Omega \Omega$, so that the dynamical phase of $S$ becomes 
the invisible axion which is a component of dark matter.  

\noindent \underline{\it Concluding remarks}~:~
In the context of $U(1)_\chi$, new light is shed on some of the outstanding 
problems in particle physics and astroparticle physics.  It is shown how 
naturally light Dirac neutrinos may be obtained in a seesaw mechanism 
which is usually reserved for considering Majorana neutrinos.  With 
light Dirac neutrinos, an elegant solution to an important problem in 
self-interacting dark matter may also be solved.  The light scalar 
mediator here is a dilepton and decays only to two neutrinos, so it 
does not disrupt the cosmic microwave background at late times. 
With the conservation of lepton number, the possibility of breaking 
baryon number $B$ to baryon parity, i.e. $(-1)^{3B}$, allows 
baryogenesis to occur, in analogy to leptogenseis, from the decay of 
a heavy Majorana fermion carrying $B=1$.  To explain strong $CP$ 
conservation, a Majorana fermion color octet with anomalous 
Peccei-Quinn charge is postulated.  It acquires a large mass through 
its coupling to a singlet scalar, the dynamical phase of which becomes 
the invisible axion and contributes to dark matter.

\noindent \underline{\it Acknowledgement}~:~
This work was supported in part by the U.~S.~Department of Energy Grant 
No. DE-SC0008541.

\bibliographystyle{unsrt}

\begin{thebibliography}{99}
\bibitem{m18} E. Ma, Phys. Rev. {\bf D98}, 091701(R) (2018) [arXiv:1809.03974].
\bibitem{m19} E. Ma, LHEP (Special Issue); arXiv:1810.06506 [hep-ph].
\bibitem{atlas-chi-17} ATLAS Collaboration, M. Aaboud {\it et al.}, JHEP 
{\bf 1710}, 182 (2017) [arXiv:1707.02424].
\bibitem{cms-chi-18} CMS Collaboration, A. M. Sirunyan, A. Tumasyan 
{\it et al.}, JHEP {\bf 1806}, 120 (2018) [arXiv:1803.06292].
\bibitem{kkm18} S. J. D. King, S. F. King, and S. Moretti, Phys. Rev. 
{\bf D97}, 115027 (2018) [arXiv:1712.01279].
\bibitem{rs84} P. Roy and O. U. Sanker, Phys. Rev. Lett. {\bf 52}, 713 (1984).
\bibitem{m02} E. Ma, Phys. Rev. Lett. {\bf 89}, 041801 (2002) 
[hep-ph/0201083].
\bibitem{mpsz15} E. Ma, N. Pollard, R. Srivastava, and M. Zakeri, Phys. 
Lett. {\bf B750}, 135 (2015) [arXiv:1507.03943].
\bibitem{m18-1} E. Ma, Mod. Phys. Lett. {\bf A33}, 1850226 (2018) 
[arXiv:1805.03295]. 
\bibitem{m15} E. Ma, Phys. Rev. Lett. {\bf 115}, 011801 (2015) 
[arXiv:1502.02200].
\bibitem{m17} E. Ma, Mod. Phys. Lett. {\bf A32}, 1730007 (2017) 
[arXiv:1702.03281].
\bibitem{kkpy17} A. Kamada, M. Kaplinghat, A. B. Pace, and H.-B. Yu, Phys. 
Rev. Lett. {\bf 119}, 111102 (2017) [arXiv:1611.02716].
\bibitem{gibm09} S. Galli, F. Iocco, G. Bertone, and A. Melchiorri, Phys. Rev. 
{\bf D80}, 023505 (2009) [arXiv:0905.0003].
\bibitem{bksw17} T. Bringmann, F. Kahlhoefer, K. Schmidt-Hoberg, and P. Walia, 
Phys. Rev. Lett. {\bf 118}, 141802 (2017) [arXiv:1612.00845].
\bibitem{m88} E. Ma, Phys. Rev. Lett. {\bf 60}, 1363 (1988).
\bibitem{fy86} M. Fukugita and T. Yanagida. Phys. Lett. {\bf B174}, 45 (1986). 
\bibitem{fps95} M. Flanz, E. A. Paschos, and U. Sarkar, Phys. Lett. 
{\bf B345}, 248 (1995) [hep-ph/9411366].
\bibitem{dm00} D. A. Demir and E. Ma, Phys. Rev. {\bf D62}, 111901(R) (2000) 
[hep-ph/0004148]. 
\bibitem{mot17} E. Ma, T. Ohata, and K. Tsumura, Phys. Rev. {\bf D96}, 
075039 (2017) [arXiv:1708.03076].
\end{thebibliography}

\end{document}